# MASSIVE NEUTRINOS AND GALAXY FORMATION

Chung-Pei Ma
Theoretical Astrophysics 130-33
California Institute of Technology, Pasadena, CA 91125
cpma@tapir.caltech.edu

ABSTRACT

We report the most recent results from high-resolution numerical simulations of structure formation in two flat cold+hot dark matter models with neutrino mass densities $\Omega_\nu = 0.2$ and 0.3. We find that structure forms too late in all CDM+HDM models with $\Omega_\nu > 0.2$ to account for the amount of dense neutral gas in high-redshift damped Lyman-$\alpha$ systems. The $\Omega_\nu = 0.2$ model at $z \approx 0$ provides a better match to observations than the pure CDM model.

## 1. $N$-BODY SIMULATIONS

Cosmological models with cold and hot dark matter (CDM+HDM) are parameterized by the neutrino mass density $\Omega_\nu$. We chose to study structure formation in two $\Omega = 1$ CDM+HDM models: (1) $\Omega_\nu = 0.3, \Omega_{\rm cdm} = 0.65, \Omega_{\rm baryon} = 0.05$; (2) $\Omega_\nu = 0.2, \Omega_{\rm cdm} = 0.75, \Omega_{\rm baryon} = 0.05$. The corresponding neutrino masses (assumed to be in one species for simplicity) are 7 and 4.7 eV, respectively. The Hubble constant is taken to be 50 km s$^{-1}$ Mpc$^{-1}$ (or $h = 0.5$). The non-linear growth of structure in the $\Omega_\nu = 0.3$ model has been shown (Davis et al. 1992; Klypin et al. 1993; Jing et al. 1994) to provide excellent fits to observations at $z = 0$. However, the worry that galactic systems form too late in this model prompted us to investigate model (2), which assumes smaller $\Omega_\nu$, causing less free-streaming suppression and earlier formation of galaxies.

We performed two high-resolution particle-particle particle-mesh (P$^3$M) $N$-body simulations to study the non-linear gravitational collapse in models (1) and (2). The same initial phases were used for the two simulations to give a controlled study of the effect of neutrino masses on galaxy formation. The primordial density perturbations were assumed Gaussian and scale-invariant. The simulation box was a cube with 100 Mpc comoving sides; the Plummer force softening was 50 kpc comoving (for $H_0 = 50$ km s$^{-1}$ Mpc$^{-1}$). A total of 23 million particles were used: 2.1 million for the cold component, and 10 times more for the hot component to sample the neutrino momenta at a given position in phase space. The particle masses in the $\Omega_\nu = 0.2$ model were $2.6 \times 10^{10} M_\odot$ for the cold and $6.6 \times 10^8 M_\odot$ for the hot component. The dynamic range and the number of particles make these the largest cosmological simulations ever published.

Potential galactic sites were identified with the Denmax algorithm described in Gelb & Bertschinger (1994) and Ma & Bertschinger (1994). In addition, an overdensity constraint of $\delta\rho/\rho \geq 200$ was imposed to eliminate the diffuse, nonvirialized dark halos.

The effect of force resolution in $N$-body simulations has been studied extensively by Gelb & Bertschinger (1994), who also demonstrated that the effect of mass resolution on the mass functions at redshifts $z \lesssim 1$ is small. Since high-redshift objects are important for constraining our models, we carried out an additional test on the effect of resolution at high redshifts. Figure 1 shows the cumulative number densities of dark halos at $z = 3$ in two $b = 2.5$ (or $\sigma_8 = 0.4$) CDM simulations with different particle masses: $2.3 \times 10^{10} M_\odot$ vs. $2.9 \times 10^9 M_\odot$. Both simulations were performed in $100^3$ Mpc$^3$ comoving boxes with $H_0 = 50$ km s$^{-1}$ Mpc$^{-1}$. The lower-resolution one (CDM16 in Gelb & Bertschinger 1994) has $144^3$ particles and a Plummer force softening of 65 kpc (for $h = 0.5$); the higher-resolution one (Jain & Bertschinger 1994) has a force softening of 20 kpc and 8 times more particles (and thus samples 8 times more wavenumbers initially). This low-amplitude CDM model produces a comparable number of halos as the $\Omega_\nu = 0.3$ CDM+HDM model at $z \sim 3$. Figure 1 shows only $< 25\%$ increase in the number densities when the mass resolution is increased by a factor of 8 and the force resolution increased by a factor $\sim 3$. Although we have not proven convergence, this test lends support to the validity of our mass resolution and high-redshift results discussed below considering that the uncertainties in cosmological observations (e.g. the normalization of the power spectrum) are generally larger than 25% and that we are ruling out models only based on order-of-magnitude discrepancies.

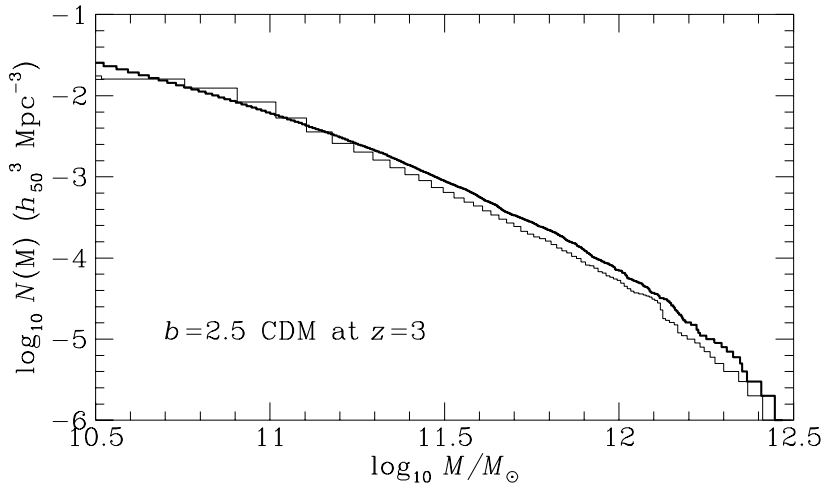

Fig. 1. The effect of resolution on the halo number densities at high redshift. The Denmax algorithm was used to identify halos; no overdensity constraint was applied. The output at $z = 3$ from two $b = 2.5$ CDM simulations is shown. The thin curve is for $144^3$ particles with particle mass $2.3 \times 10^{10} M_\odot$ and the thick curve is for $288^3$ particles with mass $2.9 \times 10^9 M_\odot$. An increase in mass resolution by a factor of 8 and force resolution by a factor of 3 makes only small difference in the number densities of halos.

## 2. HIGH-REDSHIFT CONSTRAINTS

Any hierarchical model of structure formation must account for the amount of collapsed objects observed at high redshifts. The quasar space density at redshifts up to $\sim 5$ has been used to constrain the CDM model (Efstathiou & Rees 1988; Turner 1991; Katz et al. 1994). Although the analytic predictions by Turner (1991) challenge hierarchical structure formation models, recent numerical studies (Katz et al. 1994) indicate that even the low-amplitude $\sigma_8 = 0.5$ CDM model is compatible with the known distribution of high-redshift quasars.

Since structure forms later in CDM+HDM models, a better constraint is provided by the amount of gas associated with the damped Ly$\alpha$ systems at $z \sim 3$ (Ma & Bertschinger 1994 and references therein). Lanzetta (1993) and Lanzetta, Wolfe, & Turnshek (1994) estimated that at $3 < z < 3.5$, a fraction $(2.9 \pm 0.6) \times 10^{-3} h^{-1}$ of the critical mass density is in the form of dense neutral gas in these systems. Figure 2 shows their measured value and the predictions from our simulations for the two CDM+HDM models. The normalization of the power spectra corresponds to $Q_{\rm rms-PS} = 17.6\,\mu K$ (Bennett et al. 1994). For comparison, we also plot the predicted $\Omega_g$ in the $\Omega_\nu = 0.2$ model for the higher normalization $Q_{\rm rms-PS} = 21\,\mu K$ obtained by Gorski et al. (1994). The increase of a factor $\sim 1.7$ at the low mass end is too small to save the $\Omega_\nu = 0.3$ model.

We conclude from Figure 2 that only the CDM+HDM models with $\Omega_\nu \lesssim 0.2$ can produce gravitational collapses early enough to account for the $\Omega_g$ inferred from damped Ly$\alpha$ systems at $z \sim 3$.

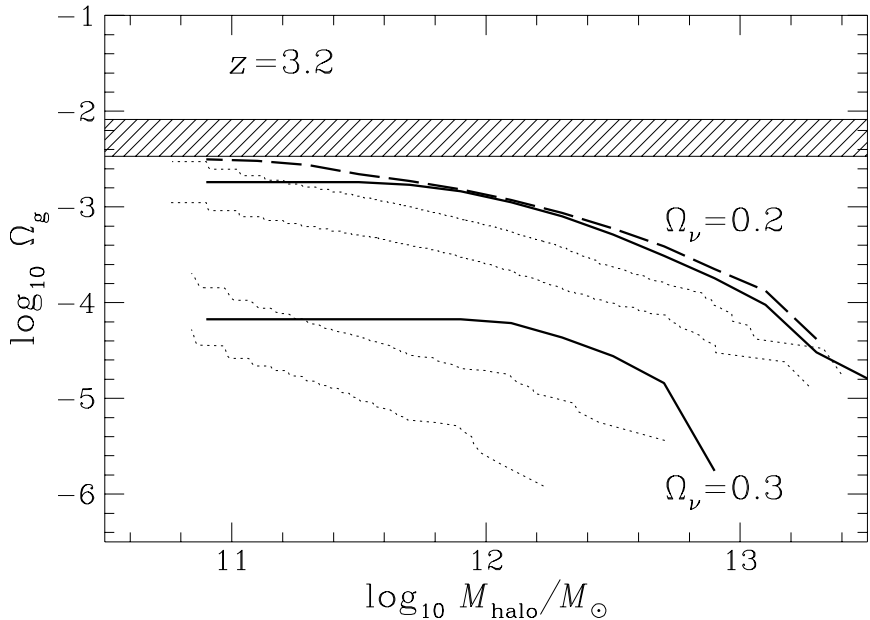

Fig. 2. The fraction of the critical density in baryons (hydrogen and helium) associated with damped Ly$\alpha$ systems at $z = 3.2$. The thick curves assume the hydrogen column density threshold of Lanzetta et al. (1993), where all halos (with no overdensity constraint) were used. The thick solid curves are for $\Omega_\nu = 0.2$ (top) and 0.3 (bottom) with $Q_{\rm rms-PS} = 17.6\,\mu K$ (bottom), and the thick dashed curve is for $\Omega_\nu = 0.2$ with $Q_{\rm rms-PS} = 21\,\mu K$ for comparison. The top two dotted curves show $\Omega_g$ predicted by the $\Omega_\nu = 0.2$ model, assuming *all* hydrogen in Denmax halos with $\delta\rho/\rho \geq 200$ (bottom) and $\geq 50$ (top) contributes to Ly$\alpha$ absorption; the bottom two dotted curves are for $\Omega_\nu = 0.3$. (All dotted curves assume $Q_{\rm rms-PS} = 17.6\,\mu K$.) The shaded band shows the observed $\Omega_g$ from Lanzetta et al.

## 3. LOW-REDSHIFT RESULTS

While the high-redshift objects discussed above provide an upper bound on the neutrino fraction $\Omega_\nu$ in CDM+HDM models, $\Omega_\nu$ is bounded from below by various $z \approx 0$ observations. After all, $\Omega_\nu = 0$ leads us back to the troubled pure CDM model. (All low-redshift results discussed below are for the $\Omega_\nu = 0.2$ model. We did not continue the simulation past $z = 1.5$ for the failed $\Omega_\nu = 0.3$ model.)

Figure 3 shows the two-point correlation function and the pairwise velocity dispersions at $z = 0.25$ (for $Q_{\rm rms-PS} = 17.6\,\mu K$). These are the two statistical quantities commonly used for comparison of models and observations. These should be compared with, e.g., Figures 8 and 12 for the pure CDM model in Gelb & Bertschinger (CDMII 1994) before their halo break-up procedures were applied. In general, the particle velocities have been reduced from 1000-1200 km s$^{-1}$ for the CDM model to 500-800 km s$^{-1}$ for the $\Omega_\nu = 0.2$ CDM+HDM, and the halo velocities have dropped from 500-600 km s$^{-1}$ to 200-400 km s$^{-1}$, in much better agreement with observations (Davis & Peebles 1983; see also Zurek et al. 1994). The correlation function for the $M \geq 2.5 \times 10^{12} M_\odot$ halos match the observed values fairly closely, although both the amplitude and the correlation length are too small for the $M \geq 2.5 \times 10^{11} M_\odot$ halos. We are continuing the simulation to $z = 0$. It remains to be seen if there will be enough evolution between $z = 0.25$ and $z = 0$ to bring $\xi(r)$ into agreement while keeping $\sigma$ reasonably low.

In Figure 3, the smaller correlation amplitude and pairwise velocity for the hot particles compared to the cold ones (dashed curves) at $r \lesssim 0.5 h^{-1}$ Mpc reflect the free-streaming effect of the neutrinos. On larger scales, the two components track each other as expected.

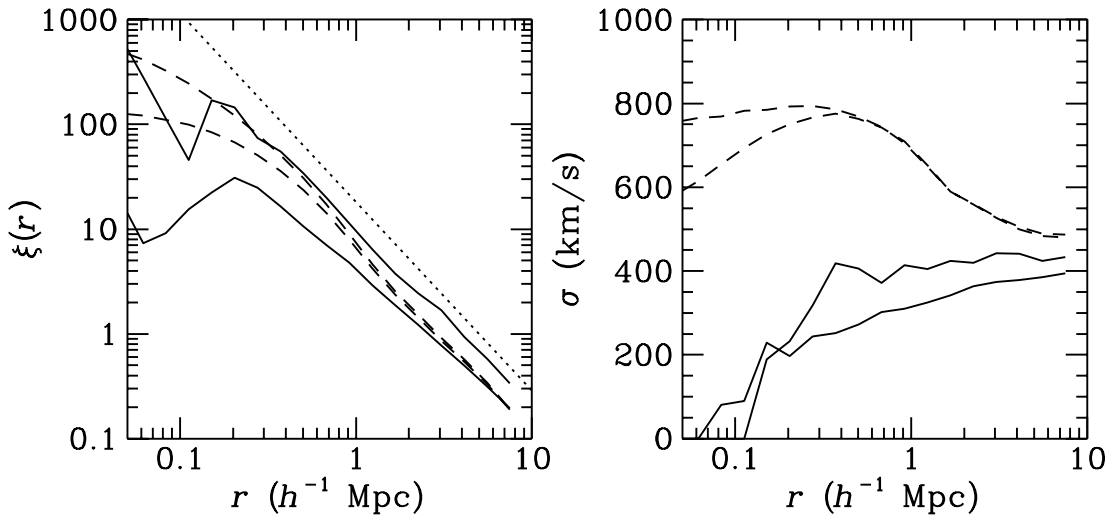

Fig. 3. The two-point correlation $\xi(r)$ and the pairwise velocity dispersion $\sigma$ at $z = 0.25$ for the $\Omega_\nu = 0.2$ CDM+HDM model. The dashed curves in both panels are for the CDM (top) and HDM (bottom) simulation particles; the solid curves are for dark halos with masses above $2.5 \times 10^{12} M_\odot$ (top) and $2.5 \times 10^{11} M_\odot$ (bottom). The dotted line indicates the observed $\xi(r) = (r/5h^{-1}{\rm Mpc})^{-1.8}$ at $z \approx 0$.

## 4. CONCLUSIONS

Between the much-studied pure CDM and HDM models span a family of mixed dark matter models which can be parameterized by $0 < \Omega_\nu < 1$, or equivalently, $0 < m_\nu < 93h^2$ eV. We are able to rule out all scale-invariant flat CDM+HDM models with $\Omega_\nu > 0.2$ and $H_0 = 50$ km s$^{-1}$ Mpc$^{-1}$ by comparing the abundance of collapsed objects at high redshifts in observations and in high-resolution $N$-body simulations. This leaves only a narrow parameter space open since very small $\Omega_\nu$ would recover the troubled CDM model. Is $\Omega_\nu \approx 0.2$ large enough to alleviate the problems encountered in the pure CDM model at low redshifts? Our simulations show that the $\Omega_\nu = 0.2$ model can better match the observed pairwise velocity dispersions and the two-point correlation function than the pure CDM model. We are analyzing other statistical quantities and completing the simulation to $z = 0$ to further scrutinize this model.

## ACKNOWLEDGEMENTS

The author acknowledges support from a PMA Division Fellowship at Caltech. The simulations were performed in collaboration with Edmund Bertschinger. Supercomputer time was provided by the National Center for Supercomputing Applications at Illinois.